\documentclass[]{article}

\usepackage{authblk}
\usepackage{graphicx}
\usepackage{color}
\usepackage{siunitx}
\usepackage{listings}
\usepackage{cite}
\usepackage{textgreek}
\usepackage{floatrow}
\usepackage{tikz}
\usepackage[scaled]{helvet}
\usepackage{xcolor}
\usepackage[draft]{todonotes}

\author[1]{Markus Lindemann\thanks{markus.lindemann@rub.de}}
\author[2]{Gaofeng Xu}
\author[3]{Tobias Pusch}
\author[3]{Rainer Michalzik}
\author[1]{Martin R. Hofmann}
\author[2]{Igor \ifmmode \check{Z}\else \v{Z}\fi{}uti\ifmmode \acute{c}\else \'{c}\fi{}}
\author[1]{Nils C. Gerhardt\thanks{nils.gerhardt@rub.de}}

\affil[1]{Ruhr-Universit\"at Bochum, Bochum, Germany}
\affil[2]{University of Buffalo, Buffalo, USA}
\affil[3]{Ulm University, Ulm, Germany}

\title{\textbf{Ultrafast spin-lasers}}

\date{}

\begin{document}
\lstset{language=Matlab}
\maketitle
The appeal of lasers can be attributed to both their ubiquitous applications and their role as model systems for elucidating nonequilibrium and cooperative phenomena\cite{DeGiorgio70}. Introducing novel concepts in lasers thus has a potential for both applied and fundamental implications\cite{Bandres18}. Here we experimentally demonstrate that the coupling between carrier spin and light polarization in common semiconductor lasers can enable room-temperature modulation frequencies above 200 GHz, exceeding by nearly an order of magnitude the best conventional semiconductor lasers. Surprisingly, this ultrafast operation relies on a short carrier spin relaxation time and a large anisotropy of the refractive index, both commonly viewed as detrimental in spintronics\cite{Zutic04} and conventional lasers\cite{Michalzik13}. Our results overcome the key speed limitations of conventional directly modulated lasers and offer a prospect for the next generation of low-energy ultrafast optical communication.

The global internet traffic will continue its dramatic increase in the near future\cite{Hecht16}. Short-range and energy-efficient optical communication networks provide most of the communication bandwidth to secure the digital revolution. Key devices for high-speed optical interconnects, in particular in server farms, are current-driven intensity-modulated vertical-cavity surface-emitting lasers (VCSELs)\cite{Michalzik13}. Analogous to a driven damped harmonic oscillator, modulated lasers have a resonance frequency $f_R$ for the relaxation oscillations of the light intensity\cite{Lee12}. For higher frequencies the response decays and reaches half of its low-frequency value at $f_\textrm{3dB}\approx\sqrt{1+\sqrt{2}}f_R$, which quantifies the usable frequency range~\cite{Michalzik13}. In conventional VCSELs the modulation bandwidth is limited by the dynamics of the coupled carrier-photon system and parasitic as well as thermal effects. The current record is $f_{3dB}=\SI{34}{GHz}$\cite{Rosales17}. Common approaches to enhance the bandwidth rely on the expression $f_R=\sqrt{v_g a S/\tau_p}/(2\pi)$, where $v_g$ is the group velocity, $a$ the differential gain, $S$ the photon density, and $\tau_p$ the photon lifetime\cite{Michalzik13}. $S$ can be increased with higher pump current and smaller mode volume, $\tau_p$ can be decreased with lower mirror reflectivities and $a$ can be optimized by material engineering. Alternative concepts to overcome the bandwidth bottleneck are actively pursued, for example, in coupled-cavity VCSEL arrays\cite{Fryslie17} or photonic crystal nanocavity lasers where $f_R>\SI{100}{GHz}$ was attained at cryogenic temperature\cite{Altug06}.

Unlike these approaches focusing on the modulation of carrier and photon densities, we consider spin-lasers which harness carrier and photon spin to improve their operation\cite{Hallstein97,Holub07,Zutic14,Chen14,Lindemann16,Torre17}.
Through the conservation of total angular momentum the spin imbalance of carriers (spin polarization) in the active region of a VCSEL is transferred to photons as circularly polarized emitted light with the polarization degree\cite{Zutic04} $P_\textrm{C}=(S^+-S^-)/(S^++S^-)$, where $S^{+(-)}$ is the right (left) circularly polarized photon density. The relation between light polarization and spin polarization of recombining carriers is given by the dipole selection rules in semiconductors.

While recent advances in spintronics show a versatile control of light polarization\cite{Nishizawa17} and THz generation\cite{Seifert16}, the ultrafast operation of spin-lasers and its fundamental limitations are largely unexplored. A tantalizing prediction of spin-lasers is that modulating the polarization of light can be realized much faster than intensity modulation (IM) and thus overcome the limitations imposed by $f_R$ of coupled carrier-photon systems\cite{Faria15}. Here, we experimentally demonstrate that adding spin-polarized carriers in conventional VCSELs supports room-temperature polarization oscillations (PO) with frequencies $>\SI{200}{GHz}$ resulting in a polarization modulation (PM) bandwidth $>\SI{240}{GHz}$, nearly an order of magnitude larger than that of the corresponding intensity dynamics. 
We show that even for high-frequency modulation there is no need for high power. Thus an order of magnitude better energy-efficiency than in the best conventional VCSELs is expected. The realization of ultrafast spin-lasers provides a different path towards emerging energy-efficient room-temperature spintronic applications, not limited by magnetoresistance\cite{Zutic04}.

\begin{figure}[!htb]
	
	\floatbox[{\capbeside\thisfloatsetup{capbesideposition={right,top},capbesidewidth=5.4cm}}]{figure}[\FBwidth]
	{\caption{Birefringent VCSEL and measurement setup. VCSEL structure with two polarization dependent refractive indices $n_x$ and $n_y$ (a). Linearly polarized electric field {\bf E}$_x$ and {\bf E}$_y$ with frequency difference $\Delta f$ (depicted duration $\ll1/\tilde{f_R}$) (b). Total electric field {\bf E} oscillating between right and left circular polarization (depicted duration $1/\tilde{f_R}$) (c). Resulting $P_\textrm{C}$ showing polarization oscillations (depicted duration $2.5/\tilde{f_R}$) (d). Schematic of experimental design with linear polarizer (LP), quarter wave plate ($\lambda/4$), lens (L) and beam splitter (BS) (e). The laser is operated with both a pumping current $J_0$ above threshold $J_{th}$ and pulsed optical spin injection. \label{fig:PO-Sektch}}}
	{
		\hspace{-5mm}
		\begin{tikzpicture}[font=\large\bfseries\sffamily]
		\node [inner sep=0pt,above right] {\includegraphics[width=6.4cm,trim=0 40 0 0, clip]{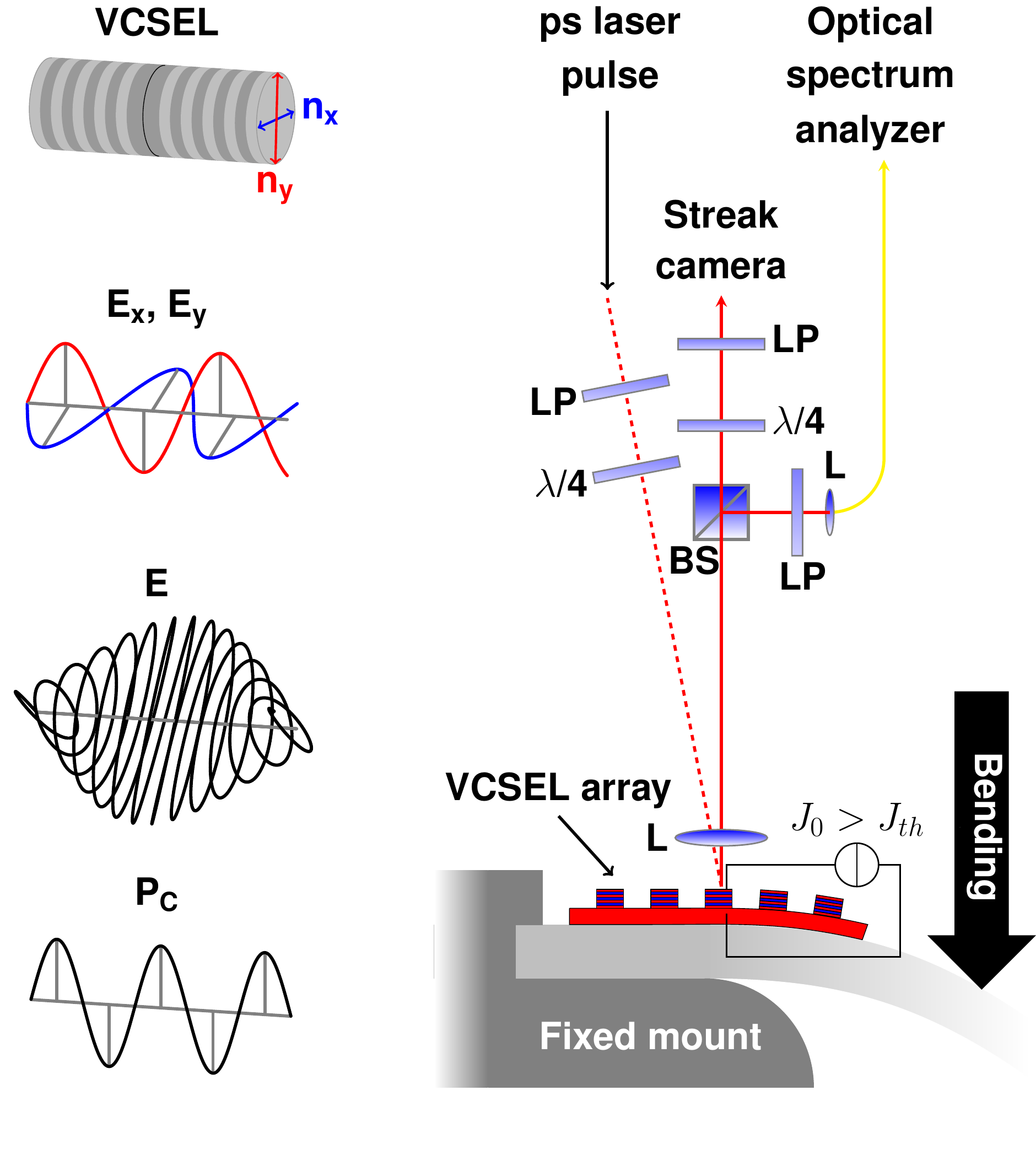}};
		\draw (0.2,6.9) node {(a)};
		\draw (0.2,5.1) node {(b)};
		\draw (0.2,3.2) node {(c)};
		\draw (0.2,1.3) node {(d)};
		\draw (3,6.9) node {(e)};
		\end{tikzpicture}
		
	}
\end{figure}
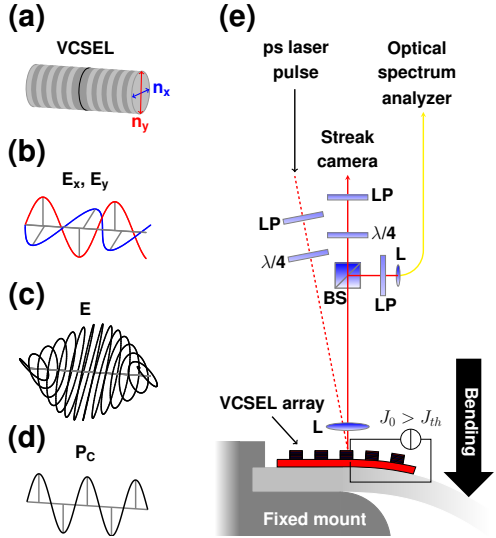

VCSELs show a linearly polarized emission due to cavity anisotropies of refractive index $n_x\neq n_y$ (birefringence, Fig.~\ref{fig:PO-Sektch}(a)) and absorption (dichroism). The orthogonal modes, described by electric fields 
{\bf E}$_x$, {\bf E}$_y$, are frequency splitted by $\Delta f$ due to the birefringence (Fig.~\ref{fig:PO-Sektch}(b)). Their usually weak coupling results in an unstable behavior with polarization switching and PO, extensively studied and attributed to residual birefringence which is considered detrimental in both conventional and spin-lasers\cite{Michalzik13,Frougier15}. In contrast, we aim to enhance the birefringence due to its connection to the PO frequency, which is related to the beat frequency between the two orthogonal modes and leads to the periodic evolution of the total {\bf E}={\bf E}$_x$+{\bf E}$_y$ (Fig.~\ref{fig:PO-Sektch}(c)). $P_\textrm{C}$ can be controlled by applying spin-polarized pumping/injection with a polarization $P_J(t)=P_0+\delta P\sin(2\pi ft)$, where $t$ is time, $P_0$ a constant offset polarization, while $\delta P$ and $f$ are amplitude and frequency of the modulation.

By generalizing\cite{Lindemann18pre} the spin-flip model for conventional VCSELs\cite{SanMiguel95}, 
our theoretical analysis reveals that the PO has a resonant behavior similar to the intensity oscillation in conventional VCSELs, but with a different frequency
\begin{equation}
\tilde{f_R}=\frac{\gamma_p}{\pi}-\frac{\gamma S_0}{4\pi(\gamma_s^2+4\gamma_p^2)\tau_p}(\alpha\gamma_s-2\gamma_p)-\frac{\epsilon_p S_0}{4\pi},
\label{Eqn:fR_tilde}
\end{equation} 
where $\gamma_p$ is the linear birefringence, $\gamma$  and $\gamma_s$ are the carrier recombination and spin-flip rates, $\alpha$ the linewidth enhancement factor, $\epsilon_p$ the  phase-related saturation and $S_0$ the steady-state photon density normalized to its value at twice the threshold 2$J_{th}$. For large $\gamma_p$ Eq.~(\ref{Eqn:fR_tilde}) holds for all practical pumping regimes, while the last two terms are negligible compared with $\gamma_p/\pi$. Thus, $\tilde{f_R}\approx \gamma_p/\pi$ suggests that strongly enhanced birefringence may overcome the frequency limitations of conventional lasers.

Among methods to enhance birefringence, for example, using anisotropic strain\cite{Faria15}, heating effects\cite{vanDoorn96,Pusch17} or photonic crystals\cite{Dems08}, we focus on mechanical bending. For this purpose we use a standard \SI{850}{nm} VCSEL\cite{Pusch16PE}, which is pumped with both a direct current above $J_{th}$ injecting spin-unpolarized carriers and a circularly polarized picosecond laser pulse exciting additional spin-polarized carriers (as an extension to purely optical spin pumping approaches, e.g. in \cite{Hallstein97, Hsu15}, Fig.~\ref{fig:PO-Sektch}(e)). The pumping conditions ensure that the spin-polarized carriers are resonantly excited in the active region to exclude influence of the split-off band\cite{Zutic04} and to minimize heating. The time- and polarization-resolved response ($f_R$ and $\tilde{f_R}$) is detected simultaneously with a Stokes polarimeter\cite{Berry77} and a streak camera, while 
$\Delta f$ 
is investigated using an optical spectrum analyzer\cite{Lindemann18pre}.
\hspace{-1cm}
\begin{figure}[!h]
\centering
			\begin{tikzpicture}[font=\large\bfseries\sffamily]
			\begin{scope}[shift={(0,0)}]
		
				\begin{scope}[shift={(-0.35,0)}]
				\node [inner sep=0pt,above right] {\includegraphics[width=6.cm]{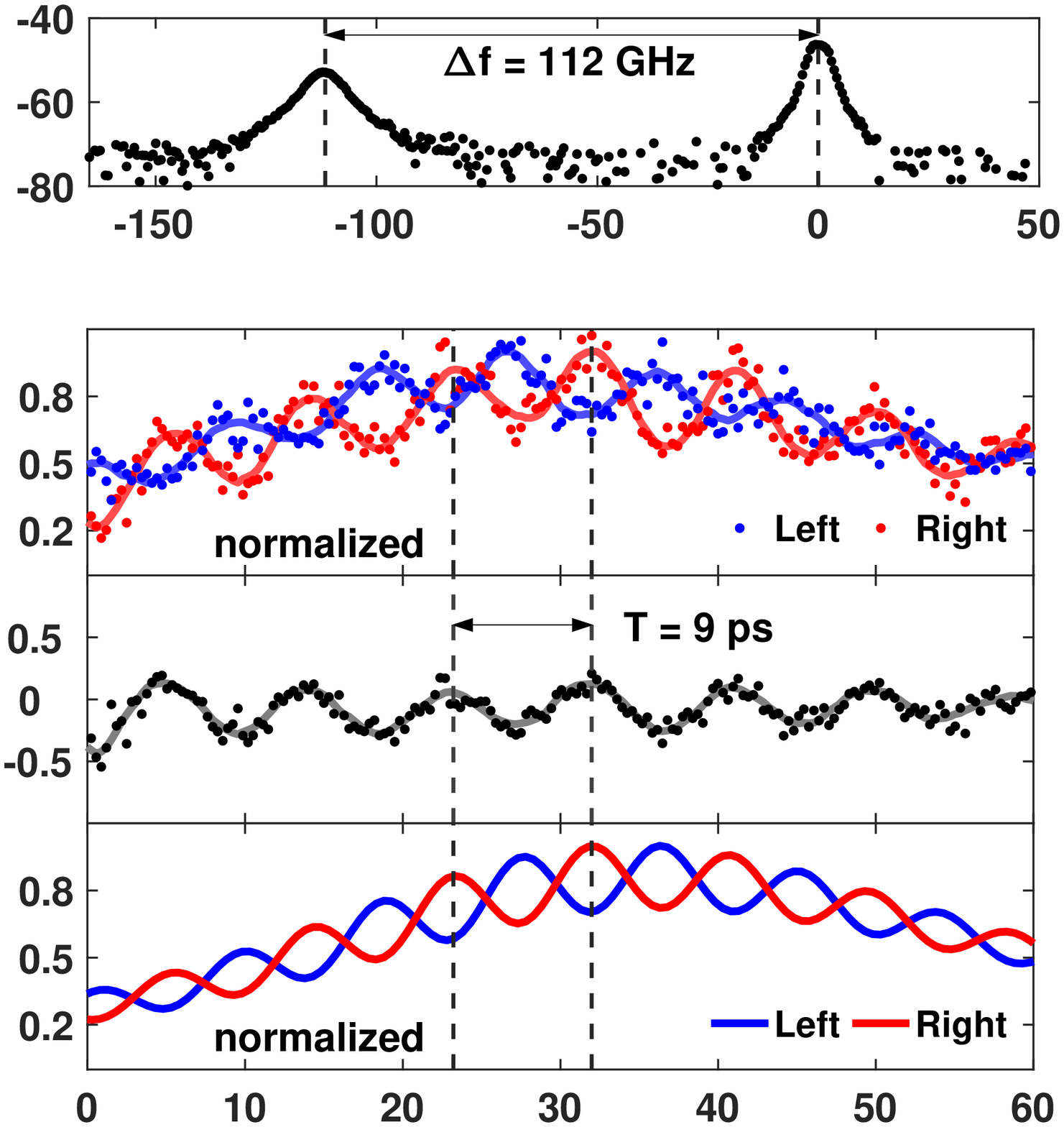}};
				\end{scope}
				\begin{scope}[shift={(5.65,-0.00)}]
				\node [inner sep=0pt,above right] {\includegraphics[width=5.95cm]{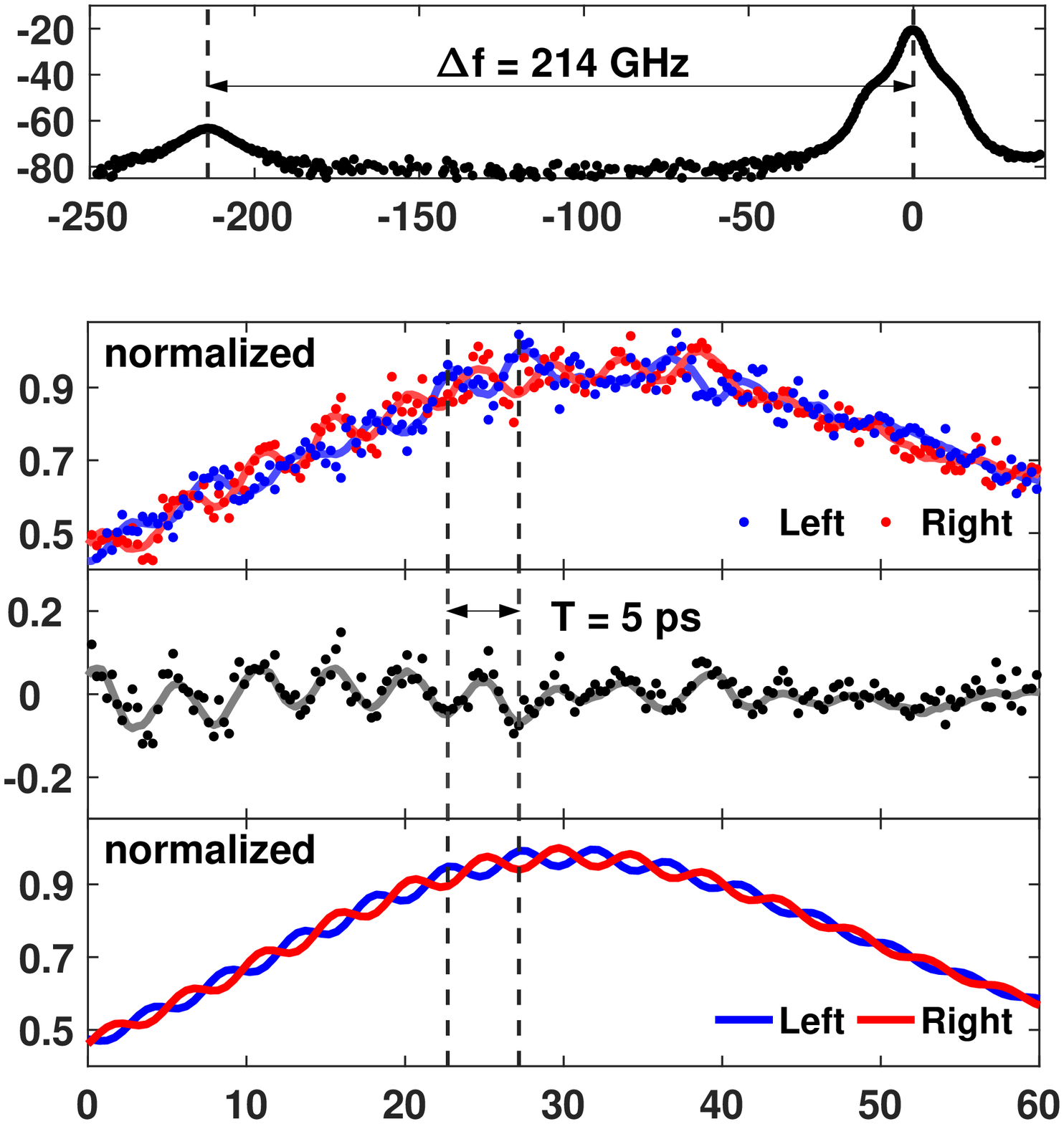}};
				\end{scope}
				\begin{scope}[shift={(-0.1,-3.8)}]
				\node [inner sep=0pt,above right] {\includegraphics[width=11.8cm,trim=26 15 0 0, clip]{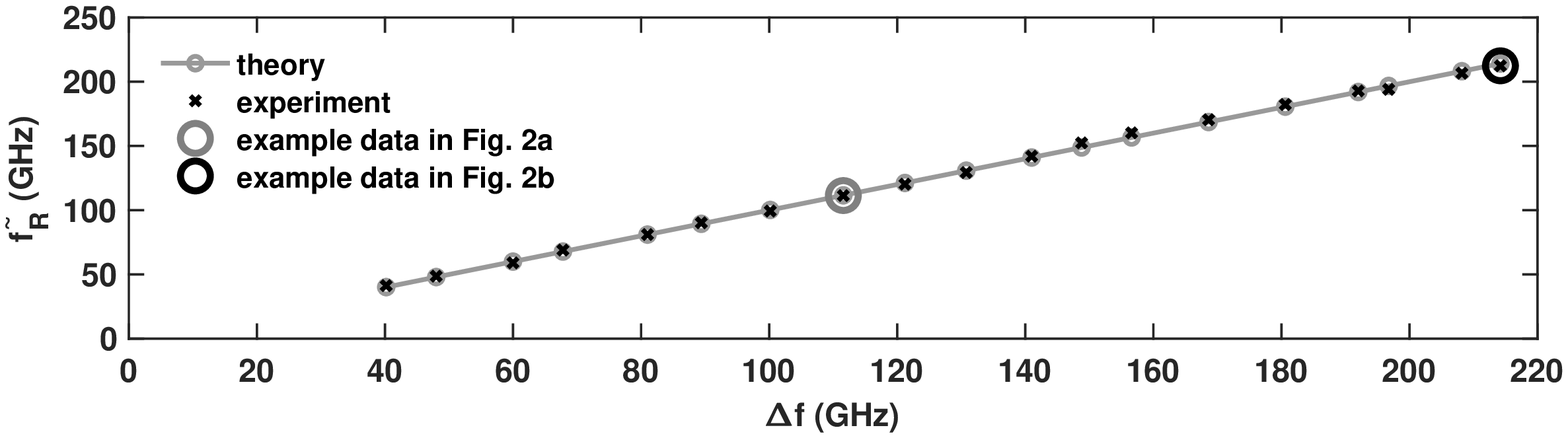}};
				\end{scope}
				\draw[black] (3,4.55) node[font=\footnotesize\bfseries\sffamily] {Relative frequency (GHz)};
				\draw[black] (9,4.55) node[font=\footnotesize\bfseries\sffamily] {Relative frequency (GHz)};
				\draw[black] (3,-.2) node[font=\footnotesize\bfseries\sffamily] {Time (ps)};
				\draw[black] (9,-.2) node[font=\footnotesize\bfseries\sffamily] {Time (ps)};
				\draw[black] (6,-4) node[font=\footnotesize\bfseries\sffamily] {$\Delta f$ (GHz)};
				\path[black] (0,1) -- node[sloped,above,font=\footnotesize\bfseries\sffamily] {S\textsubscript{Sim}} (0,1.001);
				\path[black] (-0.2,1) -- node[sloped,above,font=\footnotesize\bfseries\sffamily] {$\pm$} (-0.2,1.001);
				\path[black] (0,2.1) -- node[sloped,above,font=\footnotesize\bfseries\sffamily] {P\textsubscript{C}} (0,2.5001);
				\path[black] (0,3.6) -- node[sloped,above,font=\footnotesize\bfseries\sffamily] {S\textsubscript{Meas}} (0,3.6001);
				\path[black] (-0.2,3.48) -- node[sloped,above,font=\footnotesize\bfseries\sffamily] {$\pm$} (-0.2,3.6001);
				\path[black] (-0.1,-2.5001) -- node[sloped,above,font=\footnotesize\bfseries\sffamily] {f\textsubscript{R} (GHz)} (-0.1,-1.2);
				\path[black] (-0.36,-3.4201) -- node[sloped,above,font=\tiny\bfseries\sffamily] {$\sim$} (-0.36,-1.2);
				\path[black] (0,5) -- node[sloped,above,font=\footnotesize\bfseries\sffamily] {I (dB)} (0,6.1);
				\draw (-0.4,6.3) node {(a)};
				\draw (5.9,6.3) node {(b)};
				\draw (-0.4,-0.3) node {(c)};
			
			\end{scope}
			\end{tikzpicture}
	\caption{Polarization behaviours of spin-VCSELs. The optical spectrum $I$ reveals the birefringence-induced mode splitting to be \SI{112}{GHz} (a) and \SI{214}{GHz} (b). The main mode (at $\SI{0}{GHz}$) was suppressed for display. $S^\pm_\mathrm{Meas}$ give the measured polarization resolved normalized photon densities after pulsed spin injection (dots: raw, line: smoothed). From $S^\pm_\mathrm{Meas}$, $P_\mathrm{C}$ is determined, showing the PO. $S^\pm_\mathrm{Sim}$ give the simulated behavior\cite{Lindemann18pre}.
	$\tilde{f_R}$ versus $\Delta f$, the polarization dynamics can be tuned by birefringence-induced mode splitting (c). Data in (a) and (b) are part of the tuning series (c). At $\Delta f=\SI{214}{GHz}$ the frequency tuning is stopped to prevent mechanical sample damage.
	}
	\label{fig:PO}
\end{figure}

Results are shown for $\Delta f=\SI{112}{GHz}$ ($\Delta f=\SI{214}{GHz}$) in Fig.~\ref{fig:PO}(a) (Fig.~\ref{fig:PO}(b)). In the polarization-resolved measured and simulated normalized intensities $S^{\pm}_\mathrm{Meas}$ and $S^{\pm}_\mathrm{Sim}$, the slow envelope is the second peak of the intensity relaxation oscillation after excitation\cite{Lindemann18pre}. 
Its frequency is $f_R\approx\SI{8}{GHz}$ for both datasets. In $P_\mathrm{C}$, the overlaying intensity dynamics vanishes and only the fast PO is evident, showing $\tilde{f_R}=\SI{112}{GHz}\approx 14 f_R$ or $\tilde{f_R}=\SI{212}{GHz}\approx 27 f_R$. Remarkably, the polarization dynamics is more than an order of magnitude faster than the intensity dynamics in the same device. The PO amplitude decreases with increasing frequency, due to both the bandwidth of the measurement system\cite{Lindemann18pre} and fundamental limitations of the polarization dynamics including dichroism\cite{Lindemann18pre}, spin-flip rate, and pumping current. By changing $\Delta f$ we continuously tune $\tilde{f_R}$ up to \SI{212}{GHz}, showing an excellent agreement with the theory over the entire frequency range (Fig.~\ref{fig:PO}(c)). $\Delta f$ and $\tilde{f_R}$ coincide within the measurement accuracy\cite{Lindemann18pre}.
 
\begin{figure}
	\centering
	\begin{tikzpicture}[font=\large\bfseries\sffamily]	
		\begin{scope}[shift={(0.5,0.4)}]
		\node [inner sep=0pt,above right] {\includegraphics[width=5.4cm,trim=25 30 0 0, clip]{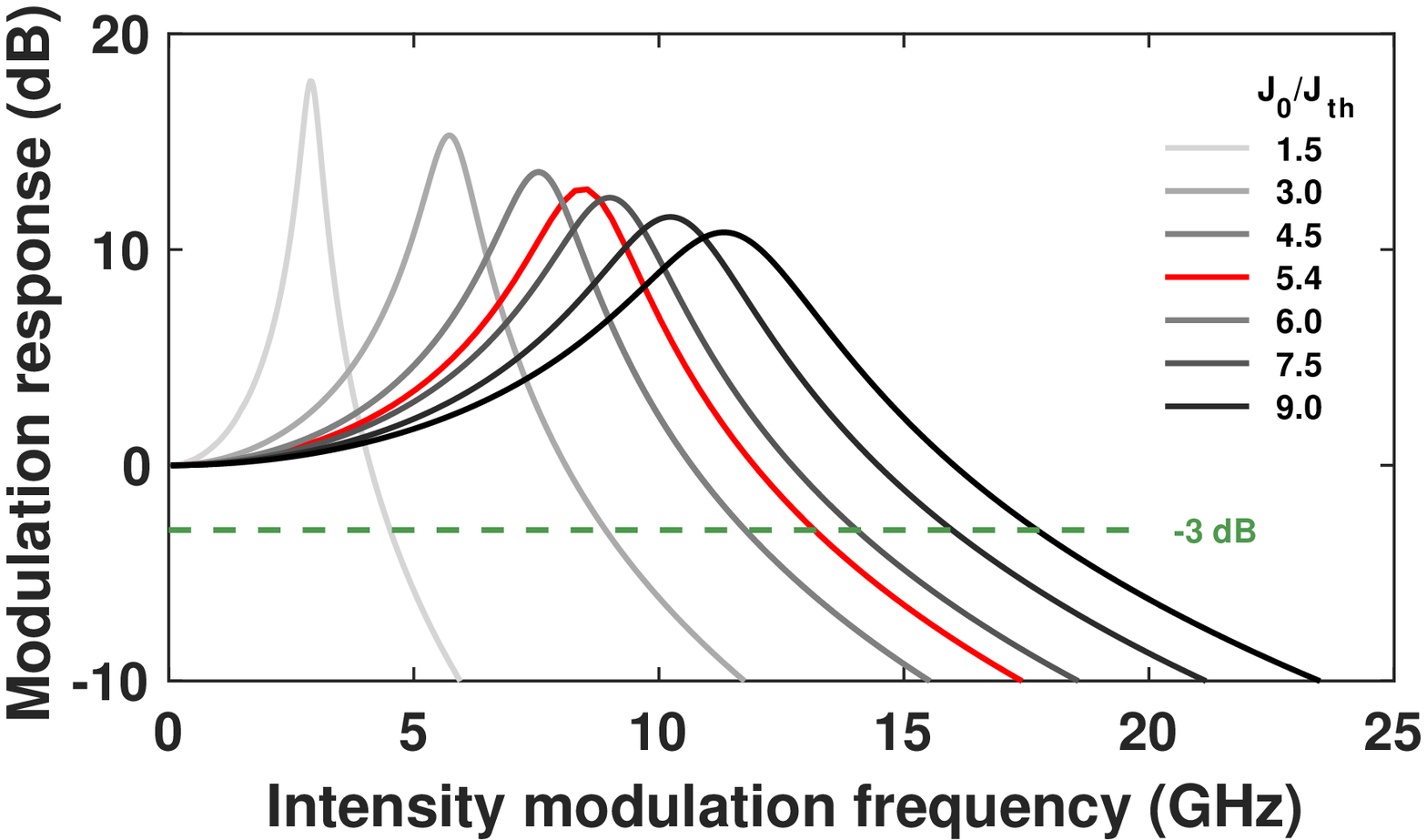}};
		\end{scope}
		\begin{scope}[shift={(6.65,0.4)}]
		\node [inner sep=0pt,above right] {\includegraphics[width=5.4cm,trim=25 30 0 0, clip]{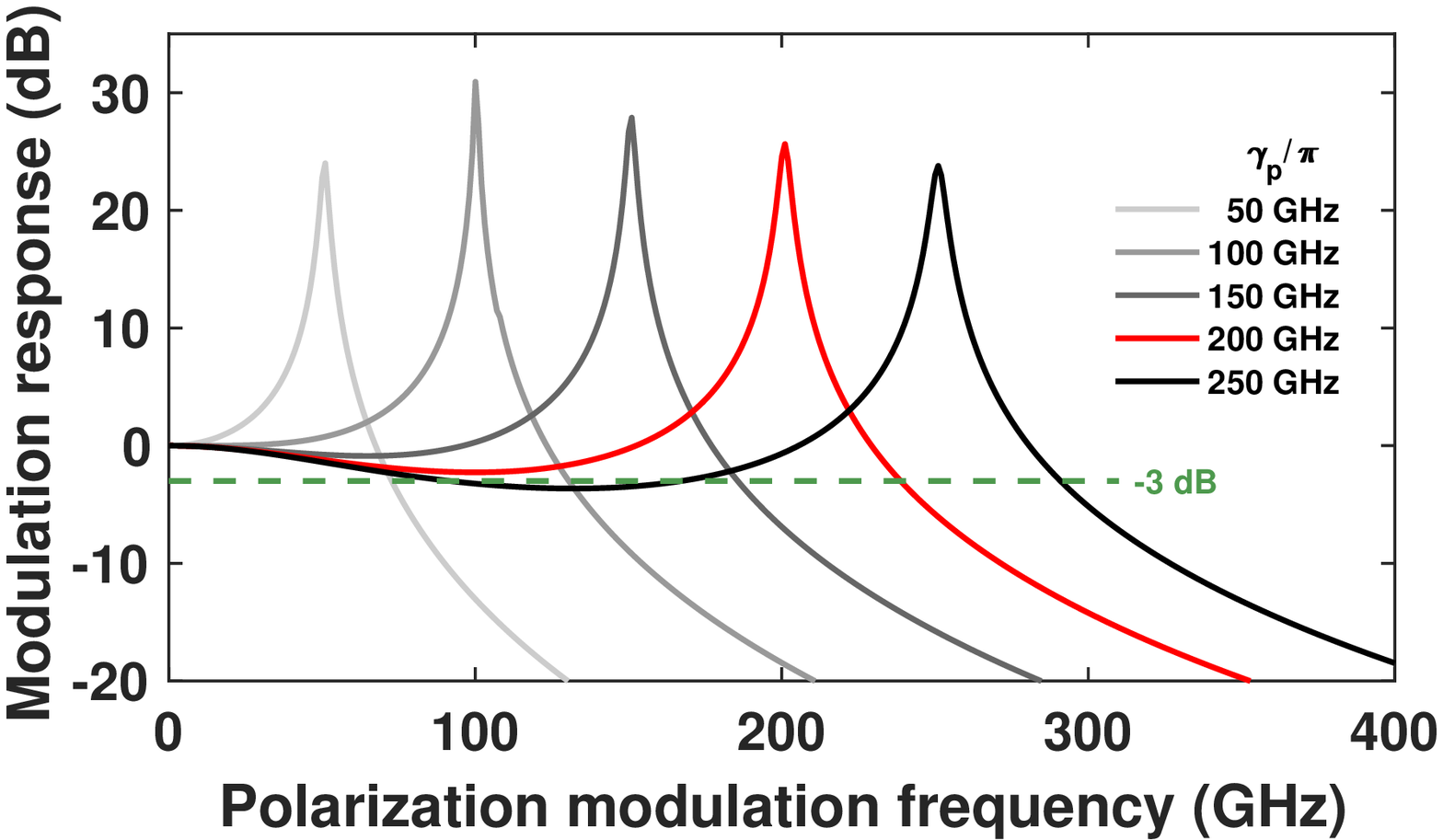}};
		\end{scope}
		\begin{scope}[shift={(0.5,-4.6)}]
		\node [inner sep=0pt,above right] {\includegraphics[width=11.6cm,trim=28 220 230 0, clip]{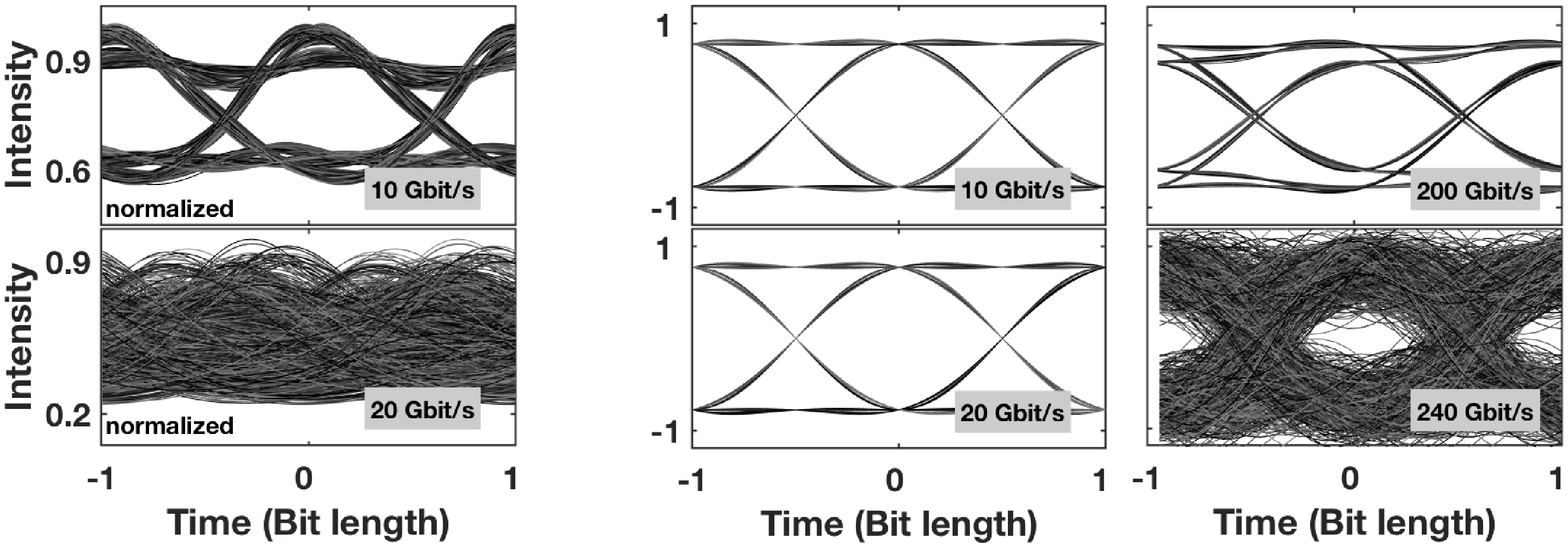}};
		\end{scope}	
		
		\draw[black] (3.3,0.15) node[font=\scriptsize\bfseries\sffamily] {Intensity modulation frequency (GHz)};
		\draw[black] (9.4,0.15) node[font=\scriptsize\bfseries\sffamily] {Polarization modulation frequency (GHz)};
		\draw[black] (2.55,-4.8) node[font=\scriptsize\bfseries\sffamily] {Time (Bit length)};
		\draw[black] (7.0,-4.8) node[font=\scriptsize\bfseries\sffamily] {Time (Bit length)};
		\draw[black] (10.35,-4.8) node[font=\scriptsize\bfseries\sffamily] {Time (Bit length)};
		\path[black] (0.55,2) -- node[sloped,above,font=\scriptsize\bfseries\sffamily] {Modulation response (dB)} (0.55,2.001);
		\path[black] (6.7,2) -- node[sloped,above,font=\scriptsize\bfseries\sffamily] {Modulation response (dB)} (6.7,2.001);
		\path[black] (0.55,-2.1) -- node[sloped,above,font=\scriptsize\bfseries\sffamily] {Intensity} (0.55,-1.401);
		\path[black] (0.55,-3.61) -- node[sloped,above,font=\scriptsize\bfseries\sffamily] {Intensity} (0.55,-3.401);
		\path[black] (5.1,-2.1) -- node[sloped,above,font=\scriptsize\bfseries\sffamily] {P\textsubscript{C} ($\%$)} (5.1,-1.401);
		\path[black] (5.1,-3.61) -- node[sloped,above,font=\scriptsize\bfseries\sffamily] {P\textsubscript{C} ($\%$)} (5.1,-3.401);
		\draw (0,3.7) node {(a)};
		\draw (6.15,3.7) node {(b)};
		\draw (0,-0.5) node {(c)};
		\draw (4.6,-0.5) node {(d)};
	\end{tikzpicture}
	\caption{Advantage of PM in dynamic performance. Simulated intensity modulation (IM), $J(t)=J_0+\delta J \sin (2\pi f t)$, response for varying normalized electric pumping $J_0/J_{th}$, where  $J_0$ is the fixed bias current and $\delta J$ is the intensity modulation amplitude (a). Polarization modulation (PM),  $P_J(t)=P_0+\delta P  \sin (2\pi f t)$, response for various birefringence (b). Red traces mark the simulations for the VCSEL under investigation at $\gamma_p/\pi=\SI{200}{GHz}$. Eye diagrams with IM using a filtered pseudorandom bit sequence for \SI{10}{Gbit/s} and \SI{20}{Gbit/s} (c). Eye diagrams with PM for \SI{10}{GBit/s}, \SI{20}{GBit/s}, \SI{200}{Gbit/s} and \SI{240}{Gbit/s} in the same device at $\gamma_p/\pi=\SI{200}{GHz}$ (d).}
	\label{fig:Datacom}
\end{figure}

The demonstrated record-high PO frequency provides the basis to evaluate the performance of spin-lasers for digital optical communication, characterized by modulation bandwidth and data transfer rates. The (IM) response in Fig.~\ref{fig:Datacom}(a) resembles the $f$-dependent displacement of a driven harmonic oscillator\cite{Lee12} with bandwidth $f_\mathrm{3dB}$, enhanced, just as $f_R$, by the increasing photon density through increased pumping. In our experiments, the pumping current of 5.4$J_{th}$ corresponds to $f_\mathrm{3dB}\approx\SI{13.5}{GHz}$, while $f_{3dB}<\SI{20}{GHz}$ for all practical currents $J_0/J_{th}$. In contrast, Fig.~\ref{fig:Datacom}(b) reveals a huge increase in PM bandwidth for the same device parameters. Similar to $\tilde{f_R}$ in Eq.~(\ref{Eqn:fR_tilde}), the corresponding bandwidth  $\tilde{f}_\mathrm{3dB}$ increases with birefringence.

To quantify the digital data transfer, a binary signal is simulated by 2$^{10}$ pseudorandom bits. The data transfer is commonly analyzed by an eye diagram\cite{Michalzik13,Wasner15} in which the time traces are superimposed. The central opening allows to distinguish between ``0" and ``1" levels. For IM, the difference in intensities of the output light is utilized. Under PM, right (left) circular polarization is encoded as ``1" (``0"). For the same set of VCSEL parameters and  $\gamma_p/\pi=\SI{200}{GHz}$, IM and PM are compared in Figs.~\ref{fig:Datacom}c,d. The closing eye diagram precludes data transfer by IM slightly above \SI{10}{Gbit/s}. PM supports the data transfer up to \SI{240}{Gbit/s}, showing a remarkable improvement in digital operation over conventional VCSELs, consistent with the increase of $\tilde{f}_\mathrm{3dB}$ over $f_\mathrm{3dB}$.
Improvements by further increasing $\Delta f$ mechanically \cite{Pusch15,Pusch17}, via photonic crystal\cite{Dems08} or strained quantum well-based VCSELs\cite{Faria15}, potentially allow $\tilde{f}_\mathrm{3dB}>\SI{1}{THz}$.

\begin{figure}[h!]
	\centering
	\begin{tikzpicture}[font=\large\bfseries\sffamily]	
	\begin{scope}[shift={(0,-0.3)}]
	\node [inner sep=0pt,above right] {\includegraphics[width=9.9cm,trim=0 0 0 0, clip]{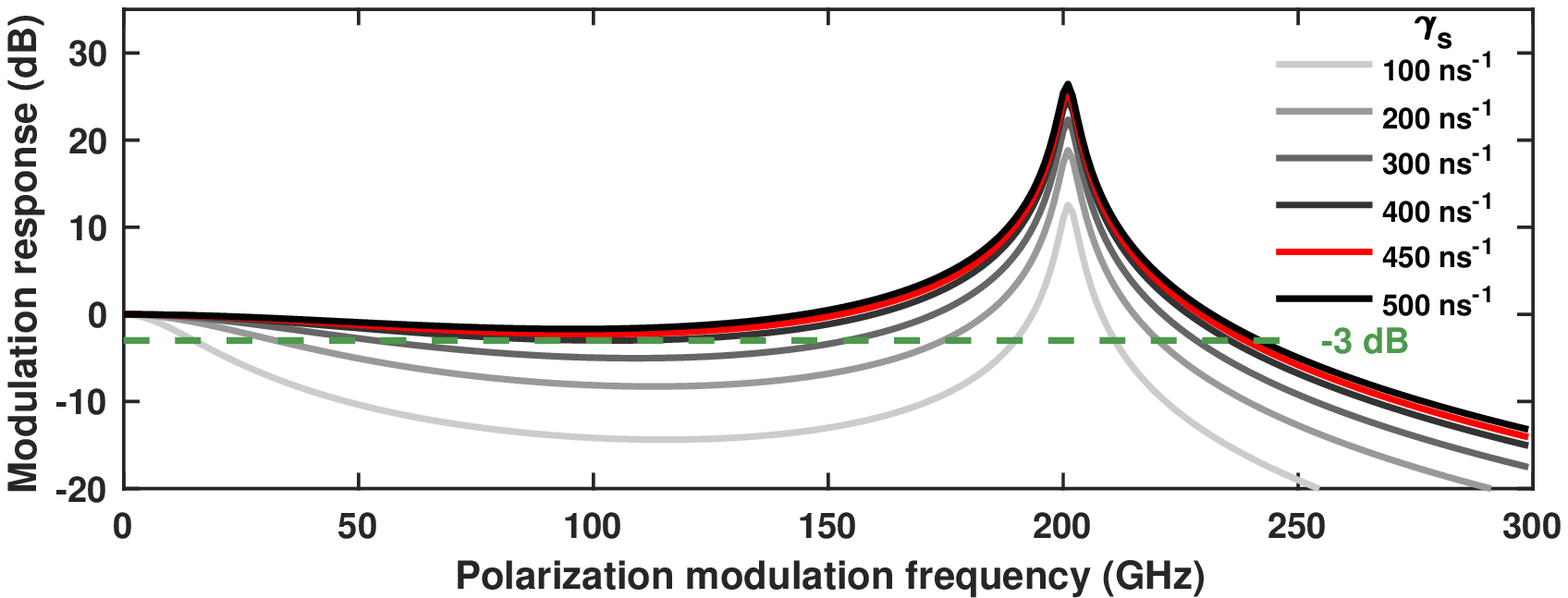}};
	\end{scope}
	\begin{scope}[shift={(0,-4.4)}]
	\node [inner sep=0pt,above right] {\includegraphics[width=9.9cm]{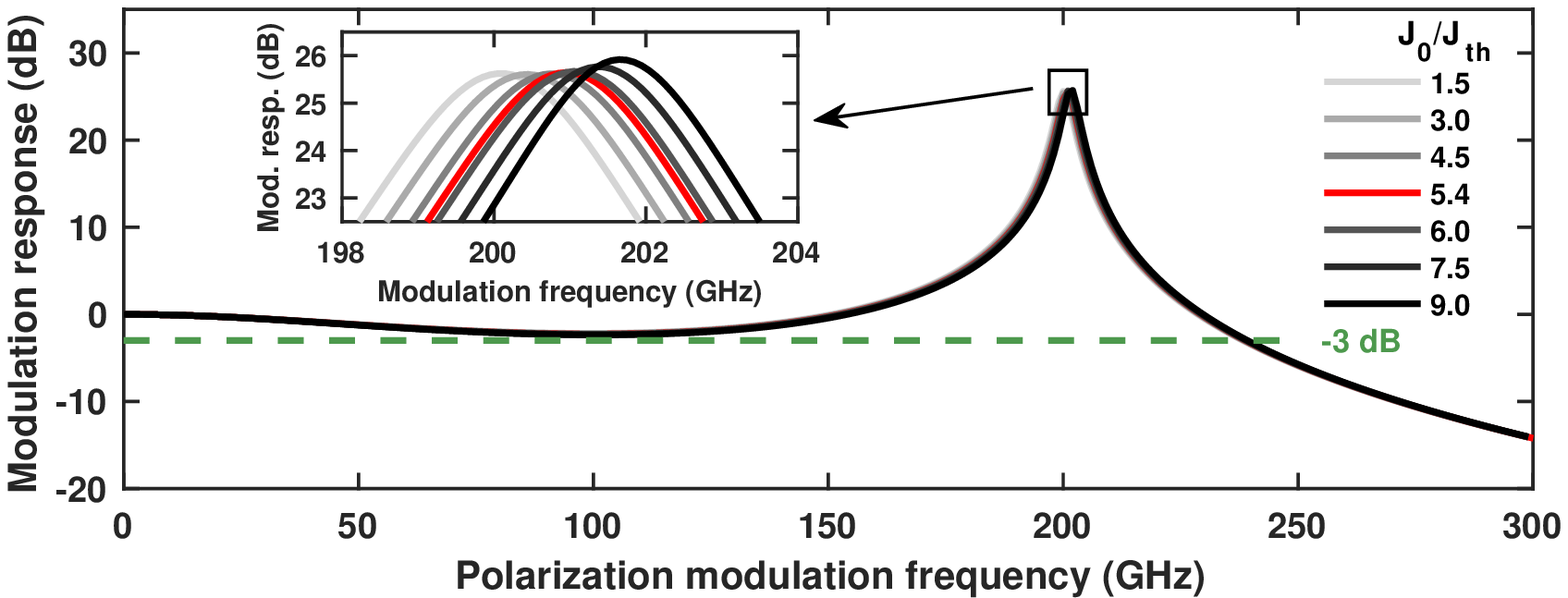}};
	\end{scope}	
	\draw (-0.3,3.6) node {(a)};
	\draw (-0.3,-0.4) node {(b)};
	\end{tikzpicture}
	\caption{Influences on modulation bandwidth. Influence of spin-flip rate $\gamma_s$ (a). Influence of normalized electric bias $J_0/J_{th}$ with the inset showing a nearly pumping-independent resonance frequency (b). Red traces mark the simulations for the VCSEL under investigation at $\gamma_p/\pi=\SI{200}{GHz}$.}
	\label{fig:MTFs}
\end{figure}

Unlike common approaches in spintronics and spin-lasers\cite{Zutic04,Iba11} which seek to increase the spin relaxation time ($1/\gamma_s$), we find that short spin relaxation times are desirable for PM. 
As long as $\gamma_s \geq 2\gamma_p/\pi$, the modulation response at $f<\tilde{f_R}$ remains above $\SI{-3}{dB}$. The simulations in Fig.~\ref{fig:MTFs}(a) reveal that in the investigated VCSEL $\gamma_s$ is close to its optimum value for $\tilde{f_R}$ slightly above \SI{200}{GHz}. $\gamma_s$ depends on the material choice\cite{Zutic04} of the active region and the VCSEL design. For example, $\gamma_s=\SI{1000}{\per\nano\second}$ was measured at room temperature for GaAs VCSELs \cite{Blansett01}, allowing for $\tilde{f_R}>\SI{500}{GHz}$. Even higher $\gamma_s$ is possible in (In,Ga)As devices.

The push for faster conventional VCSELs as well as other photonic devices typically requires a stronger pumping for IM, which leads to fundamental limitations. For example, simulating the device while neglecting heating effects, an increase in pumping from $1.5 J_{th}$ to $9 J_{th}$ enhances 
$f_{3dB}$ from $\SI{4.5}{}$ to $\SI{17.8}{GHz}$. However, higher pumping generates higher dissipated power increasing the laser temperature. In contrast, for PM $\tilde{f_R}$ is almost independent of pumping (Fig.~\ref{fig:MTFs}(b)). Thus, the highest bit rates can be already attained slightly above threshold. This enables ultra-low-power optical communication. In a conventional \SI{850}{nm} VCSEL, a heat-to-data ratio $\rm{HDR}=\SI{56}{fJ/bit}$ at $\SI{25}{Gbit/s}$ was demonstrated\cite{Moser12}. Utilizing PM for our devices, assuming pumping at $1.5 J_{th}$ with electrical spin-injection, a much lower $\rm{HDR}=\SI{3.8}{fJ/bit}$ could be obtained at a substantially higher bit rate of \SI{240}{Gbit/s}.

We have revealed a new approach to overcome the bandwidth bottleneck by utilizing polarization as the information carrier in highly birefringent spin-lasers. Even faster operation and lower energy-to-data ratios can be expected if PM is combined with high spontanous emission coupling or threshold-less nanolasers in future optical interconnects\cite{Jagsch18}. Lasers based on two-dimensional materials\cite{Wu15} which support very large strain and thus high birefringence\cite{Faria15} may offer unprecedented bandwidths. A surprising path towards desired performance relying on a high birefringence and short spin relaxation times may also stimulate other advances and unexplored phenomena in spintronics such as utilizing polarization chaos in VCSELs for secure communication and high-speed random bit generators \cite{Raddo17}. Our results show the enormous potential of spin-lasers, further motivating developing electrically pumped spin-lasers at room-temperature \cite{Saha10}.

\section*{Acknowledgements}
For supporting this work the authors thank the German Research Foundation (Grants No. GE1231/2-2 and MI607/9-2), the National Science Foundation (Grants No. ECCS-1508873 and ECCS-1810266) and the Office of Naval Research (Grant No. 000141712793).

\bibliographystyle{unsrt}

\end{document}